\documentstyle[11pt,newpasp_203_prep,twoside,epsf]{article}
\def\edcomment#1{\iffalse\marginpar{\raggedright\sl#1\/}\else\relax\fi}
\reversemarginpar

\begin{document}

\title{The formation of G-band bright points\\ I: Standard LTE modelling}
 \author{Dan Kiselman}
\affil{The Royal Swedish Academy of Sciences, Stockholm Observatory,
SE-133\,36\,\,Saltsj{\"o}baden, Sweden}

\author{Robert J. Rutten}
\affil{Sterrekundig Instituut, Postbus 80\,000, NL-3508 TA Utrecht, The Netherlands}
\author{Bertrand Plez}
\affil{GRAAL, Universit\'e de Montpellier II, FR-34095 Montpellier Cedex 5, France}

\begin{abstract}
Assuming LTE, we synthesise solar G band spectra from the
 semiempirical flux-tube model of Briand \& Solanki (1995).
The results agree with observed G-band
bright-point contrasts within the uncertainty set by
the amount of scattered light.  We find that it is the weakening of
spectral lines within the flux tube that makes the bright-point contrast
in the G band exceed the continuum contrast.
We also synthesise flux-tube spectra assuming LTE 
for the full wavelength range from UV to
IR, and identify other promising passbands for flux-tube observations.
\end{abstract}

\begin{figure}[t]
\plotone{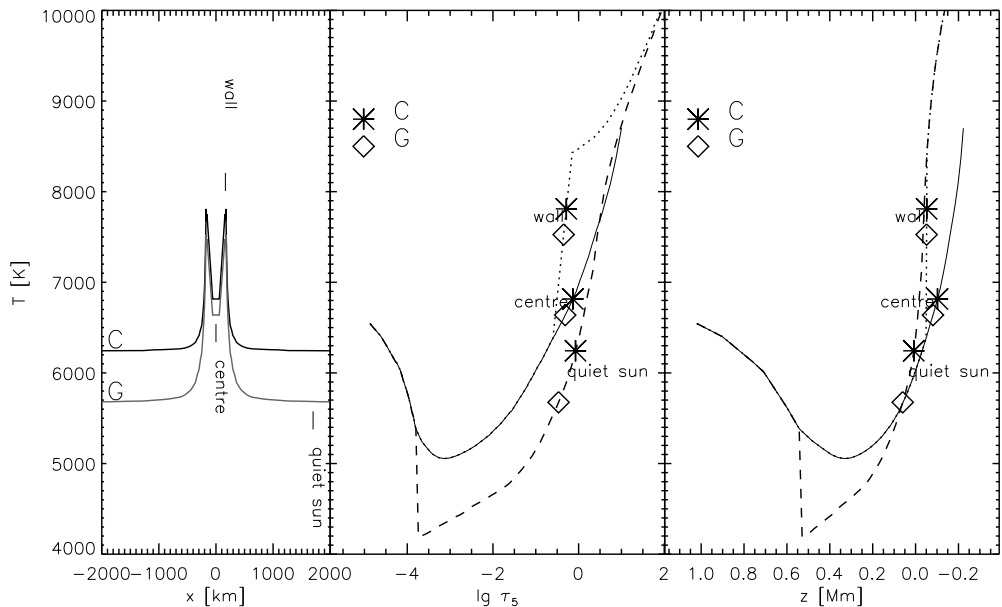}
\caption{The first panel displays emergent intensity in the 
  form of radiation temperature $T_{\rm rad}$. 
  The other panels show the kinetic temperature along the three lines of sight
  indicated in the first panel against optical depth and geometrical height,
  with the locations where $T_{\rm kin} = T_{\rm rad}$ marked.  $C$ =
  continuum, $G$ = G band.} 
  \label{allplot}
\end{figure}

\begin{figure}
\plotfiddle{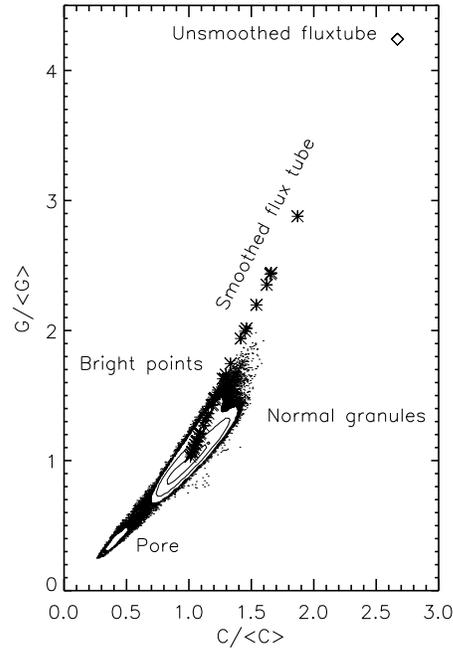}{8.5cm}{0}{70}{70}{-150}{0}
\caption{Observed normalised intensities (points and contours) and
  synthesised flux-tube 
  peak intensities for different smoothing values (stars
  and diamond). \label{fig_cgplot} }
\end{figure}

\begin{figure}
\plotone{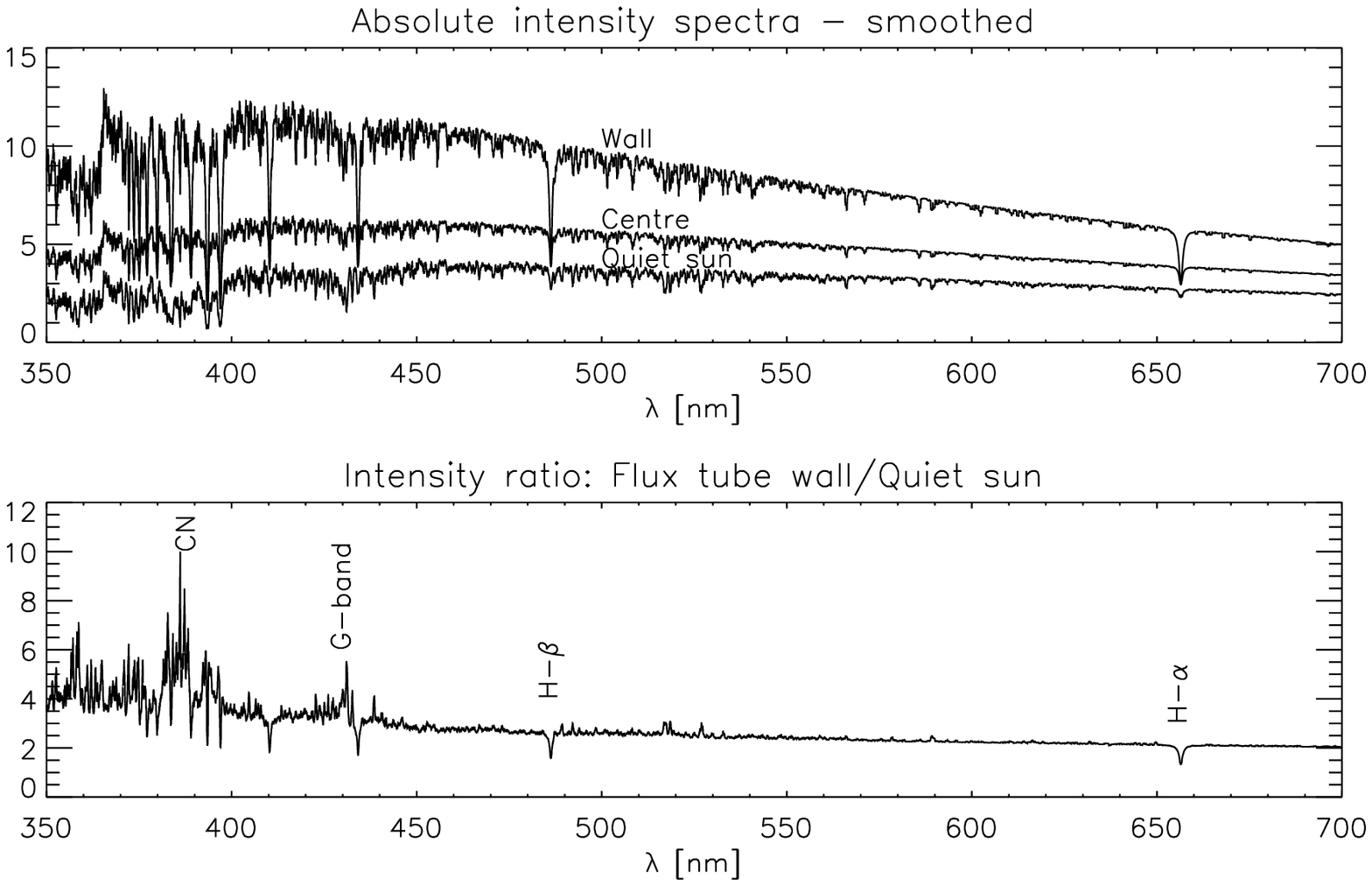}
\caption{Upper panel: entire optical spectrum from the three lines of sight 
  indicated in Fig.~1.
  Lower panel: maximum contrast
  with the outside quiet photosphere.\label{fig_specplots}}
\end{figure}

\section{Spectrum synthesis of G-band bright points}

Filtergrams taken in the G-band around 430.5~nm are often used
for proxy magnetometry since
photospheric bright points associated with magnetic structures
show up very well in them (e.g.\ Berger et al.\ 1995).
We have performed LTE spectrum synthesis of the G band (mostly formed by CH)
using the semi-empirical magnetic flux-tube model NCHROM7 of Briand \&
Solanki (1995) which is a refinement of the models of Bruls \& Solanki
(1993) and Solanki (1986). 
It is embedded in a standard model for the quiet photosphere.

Figure~1 shows results.
The hot bright walls of the (rotationally symmetric) flux tube create
a bright ring seen as a double peak in the radiation temperature.
Figure~1 
also shows the temperature structure along three lines of sight:
through quiet sun, through the tube centre, and through
the tube wall where the emergent intensity peaks considerably
because it samples the quiet-sun model at larger depth than
the quiet-sun line of sight does.
The peak contrast with respect to the quiet sun is larger in $G$
than in $C$ mainly because the photosphere appears darker in the G band.

Detailed inspection of the synthetic spectra shows how most spectral
lines weaken in the flux tube.  We interpret this as due to the
shallower temperature gradient and the lower density inside the flux
tube. Molecular lines weaken the most due to the pressure dependence
of the molecular association equilibria. This is the reason that
bright points should show up particularly well in molecular lines.

\section{Comparison with observations}

The resulting intensity spectra
were integrated using the transmission profiles
of two filters that have been used at the Swedish Vacuum Solar Telescope
(SVST)
on La Palma, denoted $G$ for the G band and $C$ for a nearby continuum band.
The results were compared to a pair of $G$ and $C$ images of a solar
active region taken with the SVST by Berger \& L{\"o}fdahl
(private communication) 
and restored by them using the phase-diversity technique
(cf. L{\"o}fdahl et al. 1998).

Figure~2 
compares the contrast values of the two
passbands -- 
i.e.\ the spectrally integrated intensity ($G$ and $C$) 
normalised to the quiet sun intensity ($\langle G\rangle$ and $\langle
C\rangle$) -- 
with the observations. The points and
contours represent the $G$ and $C$ contrasts measured per 
pixel in the SVST images. 
The stars represent peak intensities from the flux tube model
after convolution with a range of smearing functions that simulate
degradation by atmospheric seeing and by the telescope. 
The resulting peak contrast is very
sensitive to the far wings of the smearing profile -- which are very
difficult to assess.  
Nevertheless, the comparison indicates reasonable agreement between
the computed peak contrasts and the highest observed contrasts.

\section{Flux-tube signatures in other spectral bands}
Could there be other spectral regions useful for proxy magnetometry or 
flux-tube diagnostics? We synthesised also a wide range of the solar spectrum
with spectral resolution $R=20000$. 
The upper panel of Fig.~3 
shows
smoothed spectra along the three lines of sight indicated at
left in Fig.~1.  The lower panel shows the wall/quiet-sun contrast.
The Balmer lines stand out because they get stronger 
due to the high temperature of the flux-tube walls. 
The Balmer jump does not strengthen, however, so that feature is 
probably not a good flux-tube diagnostic.

The violet CN band produces large contrast in the
lower panel of Fig.~3 and is known to display the photospheric network well
(Chapman 1970). Inspection of the computed molecular equilibria shows that
CN is depleted to a smaller fraction in the flux tube than CH.
Preliminary results from the SVST
received during the symposium (Rouppe van der Voort, private
communication) confirm indeed that CN bright points
look very similar to those seen in the G band. 

\section{Conclusions}
The Briand \& Solanki model produces bright-point contrasts 
similar to those observed at the SVST.
However, the spectrum synthesis was made assuming LTE 
and the photospheric part of the flux tube model 
was primarily constructed from LTE inversions of spectropolarimetric data.
Do we believe these LTE assumptions? 
Not necessarily.
A contrasting scenario would be
one where UV radiation from the hot walls photodissociates CH in the
flux tube and also causes overionisation of metals, thus affecting
our G-band synthesis as well as the semi-empirical modelling.
If there is any place in the solar 
photosphere where 3D--NLTE effects are important, it should
be in these flux tubes.

\acknowledgements
Carine Briand and Sami Solanki very kindly provided their 
flux-tube models and software to handle them plus instructions.
Mats L{\"o}fdahl and Tom Berger generously shared their observational data.

\end{document}